\documentclass[conference]{IEEEtran}
\usepackage{comment}
\IEEEoverridecommandlockouts

\usepackage{cite}
\usepackage{amsmath,amssymb,amsfonts}
\usepackage{algorithmic}
\usepackage{graphicx}
\usepackage{tabularx}
\usepackage{textcomp}
\usepackage{xcolor}
\usepackage{url}

\begin{document}

\title{Adaptive Edge-Cloud Inference for Speech-to-Action Systems Using ASR and Large Language Models}

\author{
\IEEEauthorblockN{Mohammad Jalili Torkamani\textsuperscript{1}, Israt Zarin\textsuperscript{2}}
\IEEEauthorblockA{\textit{School of Computing, University of Nebraska--Lincoln} \\
Lincoln, Nebraska, USA \\
Email: mJaliliTorkamani2@huskers.unl.edu, izarin2@huskers.unl.edu}
}

\maketitle

\begin{abstract}
Voice-based interaction has emerged as a natural and intuitive modality for controlling IoT devices. However, speech-driven edge devices face a fundamental trade-off between cloud-based solutions, which offer stronger language understanding capabilities at the cost of latency, connectivity dependence, and privacy concerns, and edge-based solutions, which provide low latency and improved privacy but are limited by computational constraints. This paper presents ASTA, an adaptive speech-to-action solution that dynamically routes voice commands between edge and cloud inference to balance performance and system resource utilization. ASTA integrates on-device automatic speech recognition and lightweight offline language-model inference with cloud-based LLM processing, guided by real-time system metrics such as CPU workload, device temperature, and network latency. A metric-aware routing mechanism selects the inference path at runtime, while a rule-based command validation and repair component ensures successful end-to-end command execution. We implemented our solution on an NVIDIA Jetson-based edge platform and evaluated it using a diverse dataset of 80 spoken commands. Experimental results show that ASTA successfully routes all input commands for execution, achieving a balanced distribution between online and offline inference. The system attains an ASR accuracy of 62.5\% and generates executable commands without repair for only 47.5\% of inputs, highlighting the importance of the repair mechanism in improving robustness. These results suggest that adaptive edge-cloud orchestration is a viable approach for resilient and resource-aware voice-controlled IoT systems.

\end{abstract}


\section{Introduction}
\label{sec:intro}

The rapid growth of the Internet of Things (IoT) has increased the need for easy-to-use and intuitive user interfaces \cite{farooq2015review}. Among these interfaces, voice-based control is one of the most convenient methods for operating IoT devices such as smart lights, sensors, and home appliances \cite{rani2017voice}.
In many IoT systems, voice commands are processed using cloud-based systems \cite{gunawan2020development}. In this approach, audio data is sent over the internet to powerful remote servers, where speech recognition and language understanding are performed. Popular voice assistants such as Amazon Alexa and Google Home rely on cloud platforms like Amazon Web Services (AWS) to handle these tasks. Alternatively, edge-based systems process data close to where it is generated, without relying on remote cloud servers or constant internet connectivity. In voice-controlled IoT applications, edge devices such as the Jetson, Raspberry Pi, or smart gateways can perform speech recognition and command interpretation locally \cite{gondi2021performance}. This reduces dependence on the cloud and can improve response time and reliability.

Cloud-based solutions provide high accuracy in Automatic Speech Recognition (ASR) and strong Large Language Model (LLM) capabilities due to the sufficient resources they contain \cite{deuerlein2021human}. Recent work has shown that large-scale foundation models can effectively handle noisy, real-world data and improve system robustness across diverse industrial and scientific domains with minimal human intervention \cite{Mahjourian2025VLSR,jaberi2025hysim,sholehrasa2025autopk}. However, cloud-based approaches depend on stable internet connectivity, introduce higher latency, and raise privacy concerns. In contrast, edge-based systems offer better privacy and offline operation but are constrained by limited computational and energy resources, strict latency requirements, and scalability challenges \cite{safaeipour2025bayes}. These constraints can lead to slower inference and reduced performance when complex models are deployed directly on edge devices.
Previous research has explored techniques such as model compression, edge-aware optimization, inference acceleration, fixed or simplified voice commands, edge-optimized inference frameworks, and taxonomic classifications of IoT voice-control devices \cite{dantas2024comprehensive, guan2024integrated}. Most of these studies focus primarily on reducing latency or cost. However, relatively few works investigate deployable hybrid systems that can dynamically balance computation between edge and cloud resources.

To address this gap, this paper proposes an adaptively routed, voice-to-action, open-source solution \footnote{All artifacts, including the source code, dataset, and experimental results, are available on \url{https://github.com/mohammadJaliliTorkamani/ASTA}}. Our solution consists of an ASR pipeline, an LLM-based inference module, and a rule-based routing component. The router dynamically selects between local (edge) and remote (cloud) processing based on real-time system metrics such as latency, device temperature, and CPU workload. This design enables privacy-aware, low-latency, efficient, and reliable natural language control of IoT devices. This approach is valuable, as IoT systems must operate under strict constraints while maintaining both efficiency and accuracy. The proposed solution is designed for real-world applications, including smart homes and assistive technologies. Furthermore, system robustness is enhanced through command history-aware validation mechanisms. By jointly considering system metrics, adaptive routing, and command validation, our solution delivers reliable, efficient, and natural voice-based control for IoT environments.

\section{Related Work}
\label{sec:related_work}

In this section, we present a literature review of systems and methods related to adaptive voice-to-IoT research works. Our focus is on approaches that convert speech into actionable commands, balance local and remote processing, and ensure privacy, reliability, and real-time responsiveness across heterogeneous IoT environments.

\textit{\textbf{Edge-Cloud Hybrid and Adaptive Voice Systems:}}
Early efforts to bridge ASR and LLMs have sought to balance latency, accuracy, and privacy. Qin et al.\cite{cite1} introduced \textit{Tiny-Align}, a lightweight bridge between ASR and LLMs operating at the embedding level (BridgeFormer, EmbedLink). While it delivers fast and precise results on-device, its performance degrades under heavy workloads. Building upon this notion of adaptability, Shibo Y\cite{cite9} employed an NSGA-II policy to dynamically route requests between edge and cloud, optimizing quality, latency, and cost simultaneously. In a similar vein, Santosh Gondi\cite{cite7} analyzed the energy–privacy trade-offs that arise in hybrid pipelines, emphasizing the need for flexible yet efficient voice systems. Complementary to these approaches, Lukas Beno\cite{cite8} demonstrated the feasibility of containerized on-edge ASR deployments, highlighting their potential to replace cloud-centric architectures with portable, privacy-conscious setups. Hewitt and Cunningham\cite{cite4} provided a taxonomy of modern IoT voice systems (e.g., Alexa, Siri), focusing on how they balance interaction, usability, and privacy. Their findings suggest that dual-mode architectures—executing commands locally when privacy is critical, or routing to the cloud when deeper reasoning is required—are essential for next-generation adaptive systems. Building upon these insights, our work adopts an intelligent router that dynamically switches between local and cloud inference according to device load, latency, and privacy constraints.

\textit{\textbf{Lightweight and Quantized ASR for Edge Devices:}}
As computational resources on IoT nodes are limited, several studies have explored model compression and quantization to enable efficient speech recognition on small devices. Feng et al.\cite{cite3} demonstrated that quantizing ASR models to 3–8 bits can reduce memory and latency by up to tenfold with negligible accuracy loss, proving that low-bit quantized models are viable for embedded environments. In parallel, Malche et al.\cite{malche2025voice} developed an 8-bit CNN-based keyword spotting system capable of running efficiently on low-power microcontrollers such as the Arduino Nano. Similarly, Nambiappan et al.\cite{nambiappan2022edge} proposed an edge-oriented speech recognition framework for human–robot communication using predefined command templates. Together, these studies illustrate that lightweight ASR systems can operate locally with high efficiency, inspiring our integration of quantized ASR modules and wake-word detection into an adaptive IoT pipeline that scales between edge and cloud operation.

\textit{\textbf{LLM–ASR Integration and Multimodal Alignment:}}
The integration of ASR with LLMs has evolved rapidly toward unified and modular architectures. Bai et al.\cite{cite12} proposed \textit{Seed-ASR}, which directly incorporates raw speech into an LLM framework for multilingual recognition without auxiliary models. Following a similar philosophy of simplicity, Zhifu Gao\cite{cite14} showed that a compact combination of a speech encoder, projector, and chat-trained LLM can outperform more complex pipelines while maintaining clarity and efficiency. Complementary to these, Wu et al.\cite{cite15} introduced compressed acoustic features to feed speech into decoder-only LLMs, while Frank et al.\cite{cite16} used a dual-encoder mechanism to align speech and text embeddings for multilingual understanding. Leo et al. develop an offline and privacy-oriented speech-to-text model that integrates the Vosk ASR engine and small NLP microservices to have real-time corrections\cite{cite2}. Earlier work by Qin et al.\cite{cite1} and Sambal Shikhar\cite{cite10} further emphasized modular, streaming pipelines where ASR, LLM, and TTS components interact in real time. Collectively, these works support a trend toward compact, edge-first architectures that can dynamically shift workloads between local and cloud environments based on real-time context.

\textit{\textbf{Text-to-Speech and Multilingual Voice Generation:}}
Parallel progress in text-to-speech (TTS) has further enriched end-to-end voice systems. Wang et al.\cite{cite11} proposed a one-stream, zero-shot TTS architecture (\textit{BiCodec}) that unifies multi-stage tokenization into a single coherent pipeline, reducing latency while improving naturalness. Edresson’s \textit{XTTS}\cite{cite17} and \textit{SC-GlowTTS}\cite{cite19} systems advanced multilingual and zero-shot synthesis, enabling rapid speaker adaptation and voice cloning across languages with minimal data. Wei Ping’s \textit{Deep Voice 3}\cite{cite20} shifted the focus toward scalability, offering an all-convolutional attention-based design capable of production-level throughput. Together, these approaches demonstrate how fast and expressive synthesis can complement low-latency ASR and LLM components. Our project adopts similar TTS principles to deliver responsive, natural voice feedback while maintaining edge-first processing and privacy-aware routing.

\textit{\textbf{Evaluation and Benchmarks:}}
To assess performance and realism, Yiming’s \textit{VoiceBench}\cite{cite18} introduced a benchmark for testing voice assistants under diverse conditions, revealing that current models often neglect real-world variability, adaptive routing, and privacy safeguards. Meanwhile, Navonil Majumder\cite{cite13} advanced text-to-audio generation through diffusion-based learning (\textit{Tango}), aligning textual meaning with generated sound for higher conceptual fidelity. These works emphasize the importance of holistic evaluation—beyond recognition accuracy—to include adaptability, contextual understanding, and responsiveness, all of which are central to our design philosophy.

\noindent In summary, prior research has explored multiple dimensions of speech-driven systems—from efficient edge-based ASR and dynamic edge-cloud routing to unified LLM–ASR architectures and multilingual TTS synthesis. Our project consolidates these advances into an adaptive, privacy-preserving, and locally prioritized voice-IoT framework that continuously monitors device metrics to decide which inference interface is used, ensuring both responsiveness and reliability.

\section{Problem}
\label{sec:problem}

As mentioned, cloud-based systems provide strong ASR and LLM capabilities but suffer from high latency, dependence on reliable internet connectivity, and privacy concerns due to remote processing. In contrast, edge-based systems offer improved privacy and offline operation but are constrained by limited computational resources, leading to slower inference and reduced accuracy \cite{cite25}. As a result, current approaches face an inherent trade-off between performance and accuracy on one hand, and privacy, integrity, and offline availability on the other. Edge-based processing often requires high CPU/GPU utilization, which increases power consumption and accelerates device overheating. Overheating can trigger thermal throttling, further degrading performance and increasing response time. Similarly, cloud-based systems introduce network latency and API delays, which can lead to unreliable or delayed command execution. Consequently, voice-controlled IoT systems are vulnerable to performance degradation, unreliable operation, and potential privacy violations \cite{cite24}. This work aims to address these limitations by developing an adaptively routed voice-to-action solution that dynamically decides, at runtime, whether a voice command should be processed locally on the edge or remotely in the cloud. By proactively managing computational load and preventing overheating, ASTA provides efficient, reliable, and privacy-conscious natural language control for IoT devices.
\section{Solution}
\label{sec:solution}

\begin{figure*}[t]
    \centering
    \includegraphics[height=0.25\textheight]{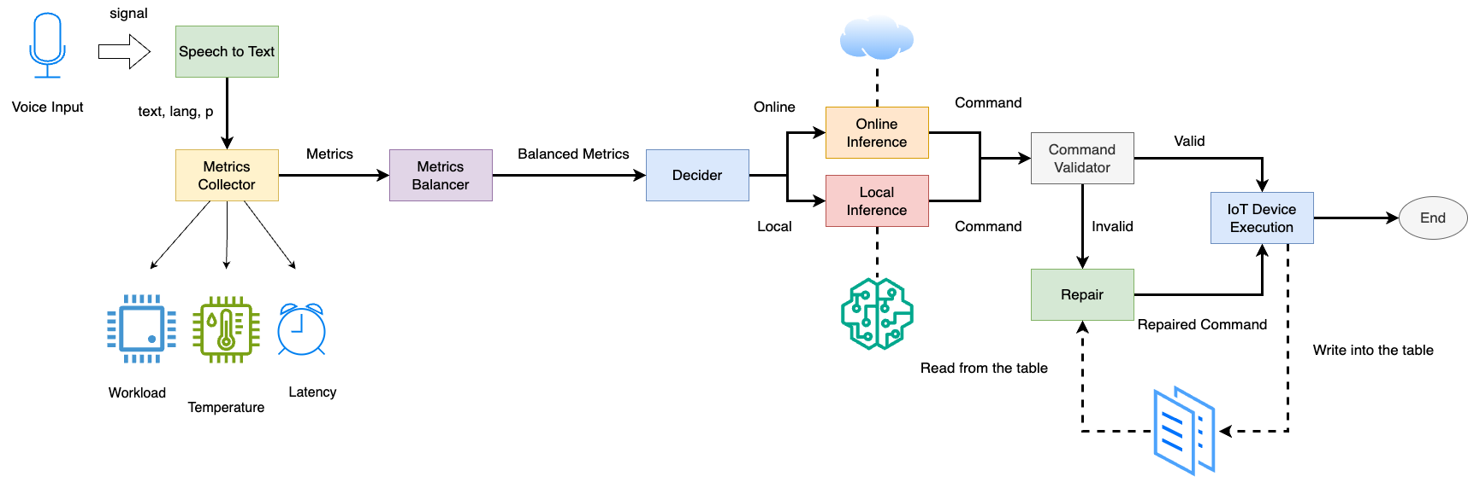}
    \caption{ASTA pipeline workflow.}
    \label{fig:workflow}
\end{figure*}

Overall, as depicted in Figure \ref{fig:workflow}, user voice input is first transcribed by an offline ASR module. Then, a metrics collection layer collects system indicators such as CPU workload, device temperature, and network latency. These metrics are evaluated by a rule-based decision-making unit, which dynamically selects between local and remote inferences. The generated commands are then validated and, if necessary, repaired using the commands history table before being executed on the target IoT devices. Finally, execution outcomes along with the user command are logged to update the user’s command history table to help potential future validations.

\subsection{Data Collection}
The workflow begins by acquiring raw audio input from a connected microphone. The audio is processed by a lightweight 8-bit tiny ASR engine based on the \textit{faster-whisper} implementation. This ASR model is optimized for fast transcription and reduced computational overhead, making it suitable for deployment on resource-constrained edge devices.

\subsection{Metrics Monitoring and Balancing}
Our tool monitors runtime system metrics, including CPU utilization, device temperature, and network latency. These metrics form the basis of the adaptive routing mechanism, enabling the system to select the most suitable inference mode at runtime. Offline inference is selected under two conditions: high CPU utilization (above 80\%) and elevated device temperature (above 50$^\circ\mathrm{C}$). This combination indicates that the edge device is under heavy computational and thermal stress, making additional network-based processing undesirable for maintaining system robustness. In addition, when network latency to the OpenAI server exceeds 150ms, online inference may fail to meet real-time responsiveness requirements. In such cases, inference is performed offline using the local LLM. In all other scenarios, online inference is preferred, as the device has sufficient available resources to establish a stable connection to cloud-based LLM APIs while maintaining acceptable latency and performance. In our experiments, since the edge device is typically underutilized, we introduce controlled, probabilistic perturbations to the monitored metrics that, with a probability of 0.5, deliberately push them beyond predefined thresholds. This mechanism ensures a balanced distribution between online and offline inference decisions.

\subsection{Online and Offline Interface}
For online inference, we used the cost-efficient yet relatively powerful GPT-3.5-turbo model via web APIs, selected for its favorable trade-off between cost and performance, while the TinyLlama-1.1B-chat-v1.0 model was employed for offline inference. This enables operation without internet connectivity while preserving user privacy. Offline inference is selected when system metrics do not remain within safe operating ranges, allowing commands to be processed locally with minimal latency, CPU load, and thermal impact.

\subsection{Command Validation}
After the LLM produces a structured command, the system performs multi-layer command validation to ensure validity, completeness, and executability. Validation is conducted across three layers: action, device, and index.

The action layer verifies that the requested operation is included and supported (e.g., turning a device on or off). The device layer checks whether the referenced device exists and is correctly identified. The index layer ensures that a specific device instance is specified. If any component is missing or invalid, the system attempts automatic command repair. For example, if a user issues the command ``turn on the light'' without specifying which light, the system selects the most frequently used light whose action is `turn on' from the history table and reconstructs the command as ``turn on light 1''.

\subsection{Execution Layer}
Once a command is validated, it is passed to the execution layer, where it is translated into a concrete IoT control action. This layer interfaces with connected devices, such as smart lights or speakers, using predefined communication protocols. Based on the validated action, device type, and index, the execution module dispatches the appropriate command to the target device. All executed actions are recorded in the history table, enabling future command inference and contextual understanding. The execution layer completes the voice-to-action pipeline by ensuring reliable, accurate, and end-to-end control of IoT devices.
\section{Evaluation}
\label{sec:evaluation}

\begin{figure}[h]
    \centering
    \includegraphics[width=0.5\columnwidth]{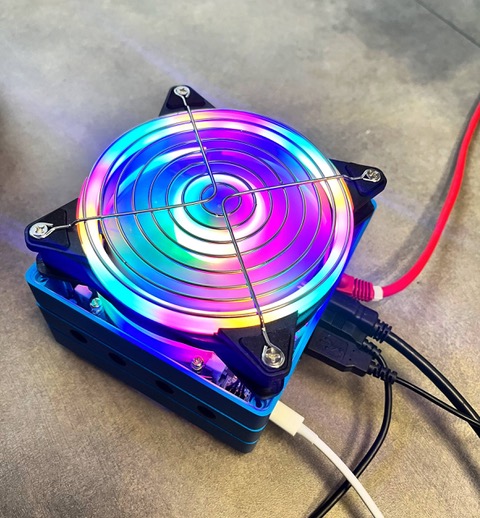}
    \caption{Jetson Mate Xavier NX edge device used for local ASR and LLM inference.}
    \label{fig:jetson_mate}
\end{figure}

\begin{figure}[h]
    \centering
    \includegraphics[width=0.5\columnwidth]{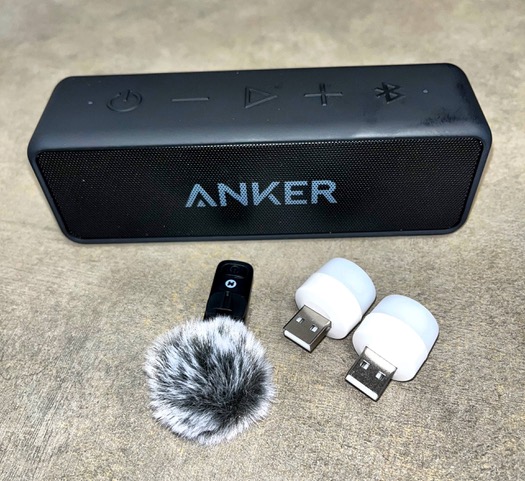}
    \caption{Accessories used in the experiment, including a microphone, speaker, and USB-powered lights.}
    \label{fig:accessories}
\end{figure}

We deployed the proposed solution on a Jetson Mate Xavier NX edge device (Figure~\ref{fig:jetson_mate}), equipped with a microphone, two USB-powered lights, and a speaker (Figure~\ref{fig:accessories}). To evaluate the system, we constructed a dataset consisting of 80 audio files, each containing an English-spoken command. The dataset includes a diverse set of prompts, such as syntactically correct and complete commands, index-variant commands, compound commands, commands involving unsupported devices, and ambiguous or irrelevant inputs. This diversity enables a comprehensive evaluation of the system’s ability to handle invalid inputs, assess ASR performance, and evaluate end-to-end command execution. Each audio file in the dataset was processed by the pipeline, and evaluation statistics were extracted from the JSON artifacts generated by the tool. 

\begin{figure}[h]
    \centering
    \includegraphics[width=0.8\columnwidth]{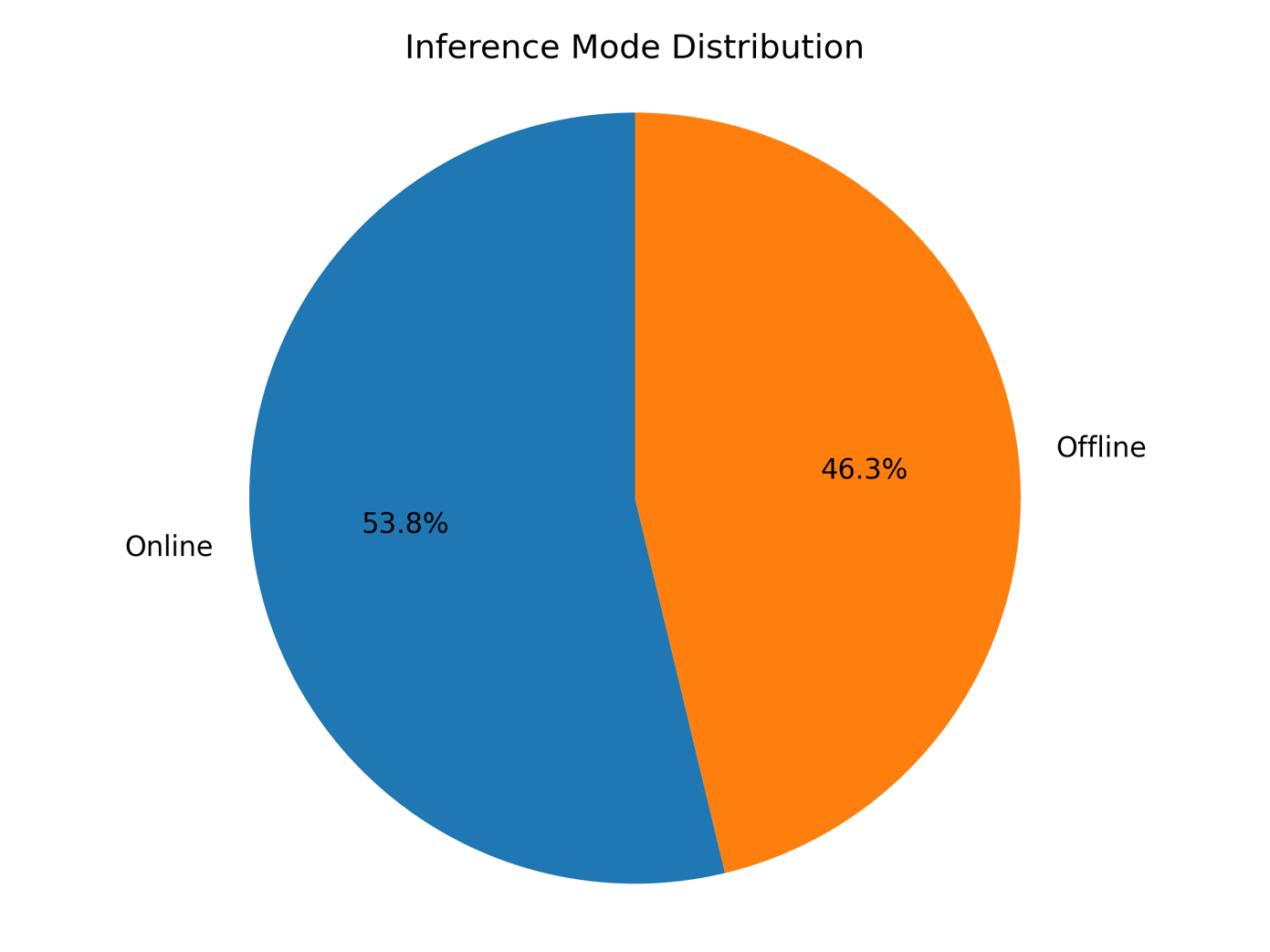}
    \caption{Distribution of inference mode decisions: online vs. offline.}
    \label{fig:distribution}
\end{figure}

Overall, our experimental results indicate that the ASTA pipeline successfully routed all input prompts to the command execution unit. As shown in Figure~\ref{fig:distribution}, 43 samples (53.8\%) were routed to online inference, while 37 samples (46.2\%) were processed offline. This near-even split is expected and primarily reflects the metric balancer factor of 50\%. Regarding ASR accuracy, 50 samples (62.5\%) were correctly transcribed, and our manual analysis shows that the most transcription errors were attributable to numeric representations (e.g., ``5'' instead of ``five'') and phonetically similar words (e.g., ``too'' instead of ``two'', or ``of'' instead of ``off''). Regarding command generation without repair, we found that the tool generated correct commands for 38 samples (47.5\%) regardless of the inference path. More specifically, 31 samples (72.0\%) were correctly generated using online inference, while the remaining were done offline. This number further indicates the superiority of the online inference model, as it is a more powerful model in terms of the number of neural parameters compared to the offline inference model. Overall, the tool correctly transcribed and generated execution commands for 23 samples (28.7\%) without requiring any repair, indicating the importance of the repairing component within our pipeline.

\begin{figure}[h]
    \centering
    \includegraphics[width=0.8\columnwidth]{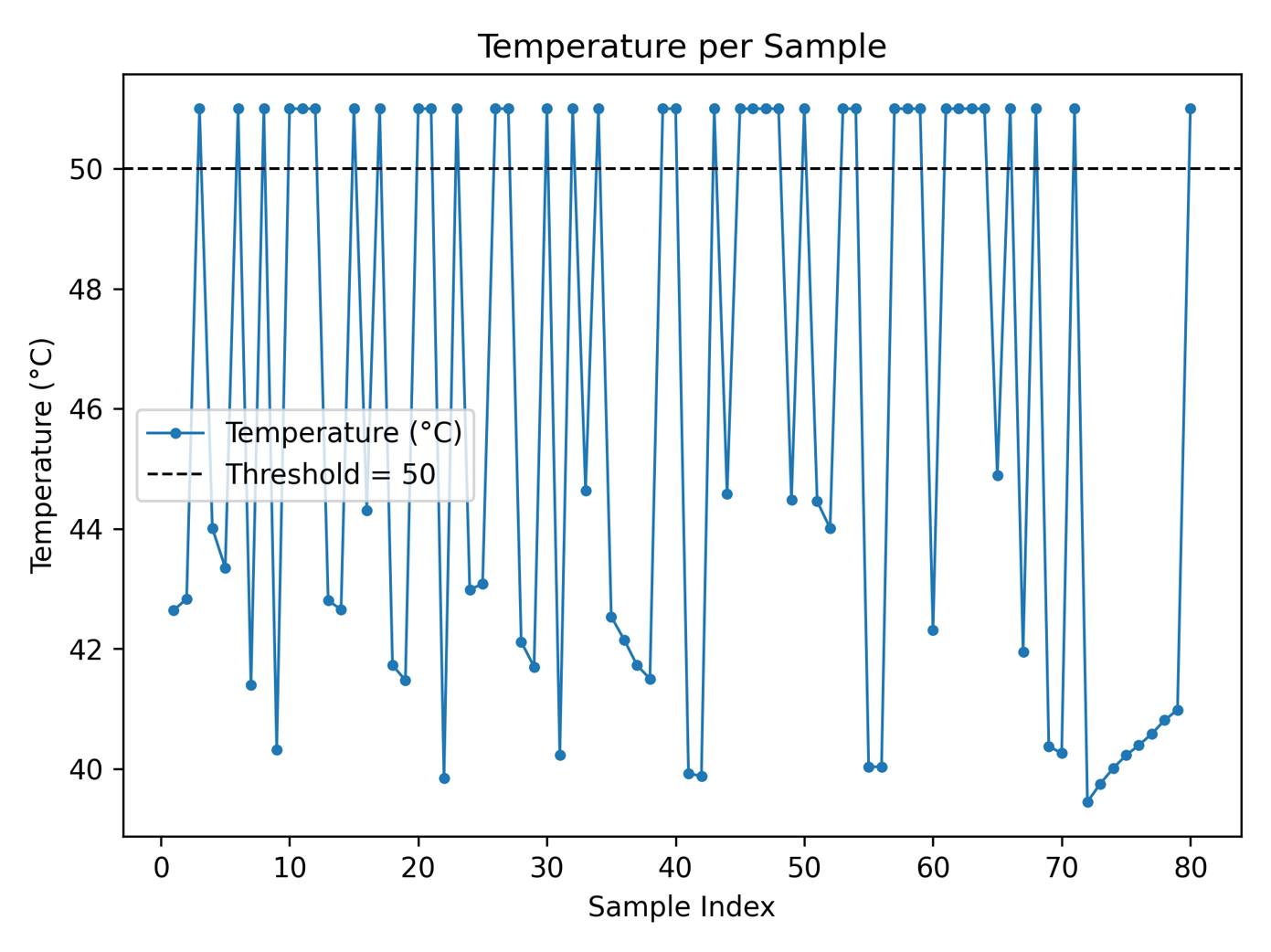}
    \caption{Average temperature of CPU cores per sample.}
    \label{fig:cpu_temperature}
\end{figure}

\begin{figure}[h]
    \centering
    \includegraphics[width=0.8\columnwidth]{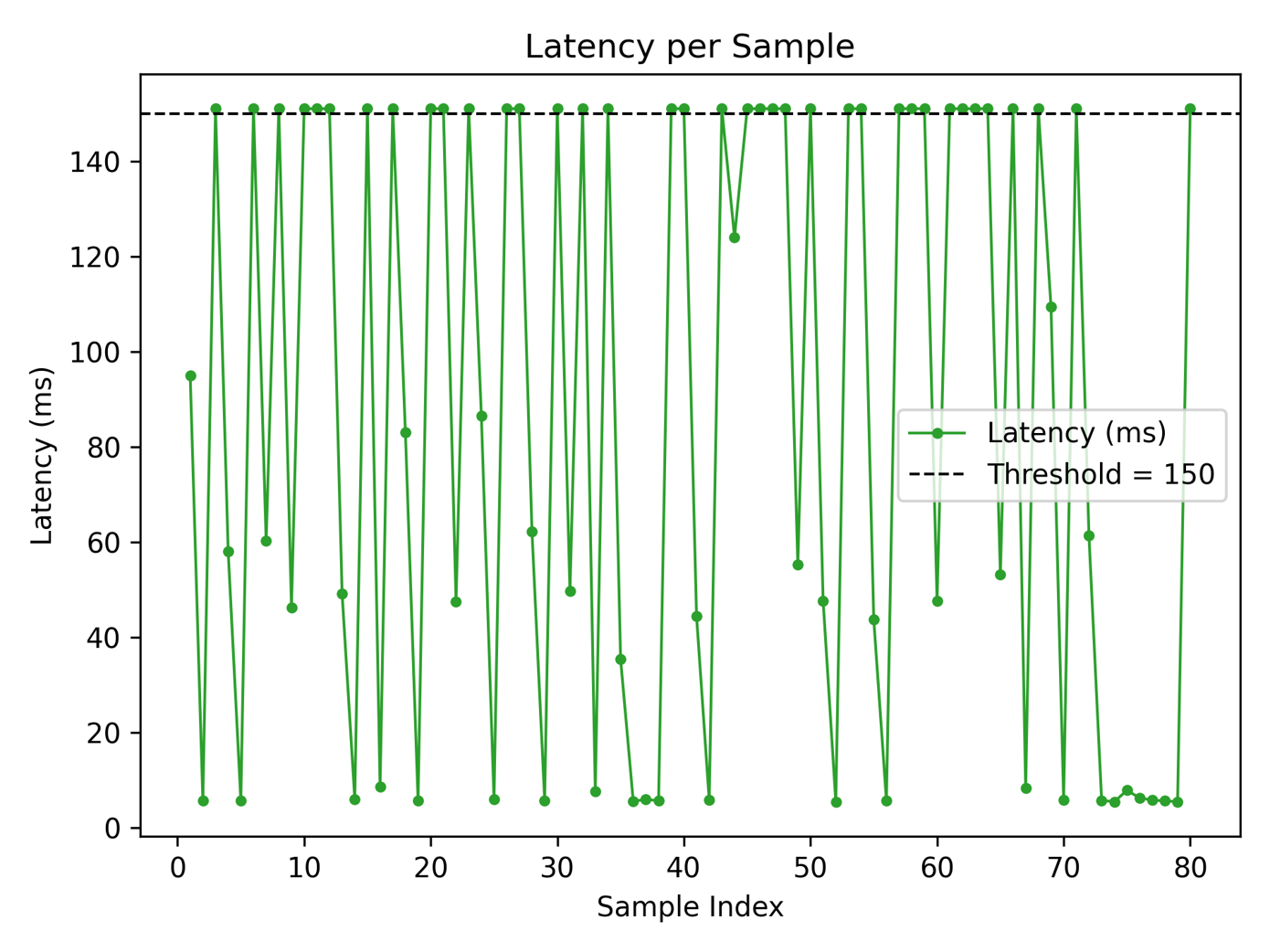}
    \caption{OpenAI inference latency per sample.}
    \label{fig:latency}
\end{figure}

\begin{figure}[h]
    \centering
    \includegraphics[width=0.8\columnwidth]{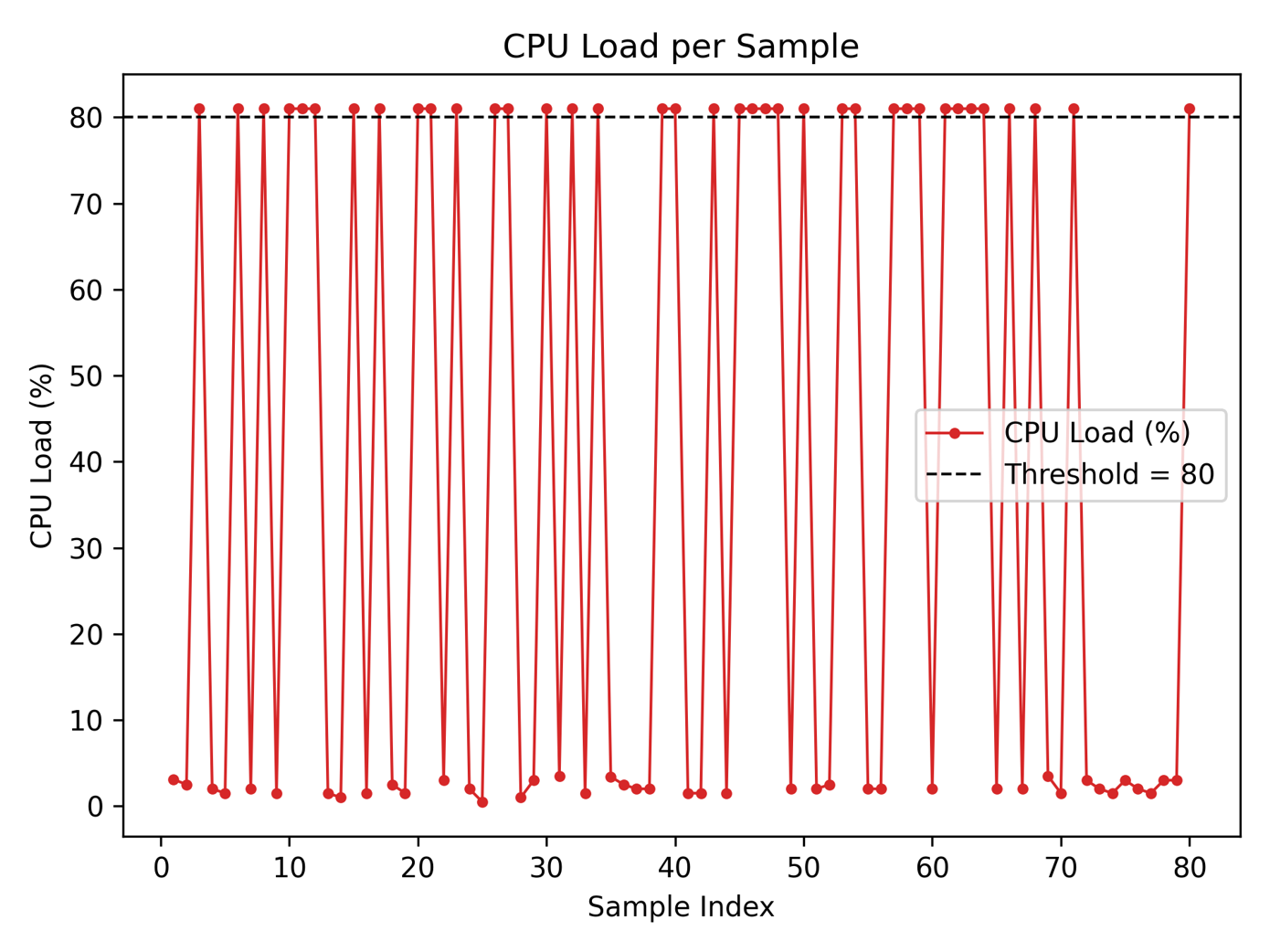}
    \caption{Average workload of CPU cores per sample.}
    \label{fig:cpu_workload}
\end{figure}

With respect to performance metrics, the system exhibited an average CPU load of 38.5\%, an average inference latency of 87.3ms, and an average operating temperature of 46.0$^\circ\mathrm{C}$, all of which fall within safe and stable operating ranges. However, these metrics should be interpreted with caution, as they are influenced by metric balancer offsets and do not exclusively reflect the intrinsic CPU workload, temperature, or network latency of the edge device. Figure~\ref{fig:cpu_temperature} illustrates the per-sample CPU core temperatures, which consistently remained within safe operating limits. Similarly, Figures~\ref{fig:latency} and~\ref{fig:cpu_workload} present the observed network latency and CPU workload during the experiments. Together, these results provide insights into appropriate threshold selection for different edge device deployment scenarios. A summary of the evaluation results is presented in Table~\ref{tab:results-summary}.

\begin{table}[h]
\centering
\caption{Summary of Experimental Results}
\label{tab:results-summary}
\renewcommand{\arraystretch}{1}
\begin{tabularx}{0.97\columnwidth}{|c|>{\centering\arraybackslash}X|}
\hline
\textbf{Metric} & \textbf{Value} \\
\hline
Total samples & 80 \\
\hline
Inference routing & 100\% (Online: 43, Offline: 37) \\
\hline
Correct ASR & 62.5\% \\
\hline
Correct command generation & 47.5\% (Online: 72\%, Offline: 18.9\%)\\
\hline
Avg. CPU workload & 38.5\% \\
\hline
Avg. latency & 87.3 ms \\
\hline
Avg. temperature & 46.0$^\circ$C \\
\hline
\end{tabularx}
\end{table}

\section{Conclusions and future work}
\label{sec:conclusion}

This paper presents an adaptive dual-path voice-to-action solution capable of dynamically switching between online and offline inference systems based on real-time system indicators. Our proposed solution, ASTA, integrates metric-aware inference routing with a rule-based prompt validation and repair mechanism prior to command execution. Experimental results demonstrate that ASTA effectively balances online and offline inference, achieves an ASR accuracy of 62.5\%, and successfully executes all input commands, thereby providing integrity and reliability for IoT and edge-AI applications. As future work, we suggest investigating the use of natural language processing techniques to identify and correct missing words during transcription and further improving the ASR accuracy.

\bibliographystyle{IEEEtran}

\bibliography{references}

\end{document}